# Software Engineering Education Beyond the Technical

## A Systematic Literature Review

**Wouter Groeneveld**
OVI and LESEC, Department of Computer Science, KU Leuven
wouter.groeneveld@kuleuven.be
https://orcid.org/0000-0001-5099-7177

**Joost Vennekens**
OVI, Department of Computer Science, KU Leuven
joost.vennekens@kuleuven.be

**Kris Aerts**
OVI, Department of Computer Science, KU Leuven
kris.aerts@kuleuven.be

## ABSTRACT

Higher education provides a solid theoretical and practical, but mostly technical, background for the aspiring software developer. Research, however, has shown that graduates still fall short of the expectations of industry. These deficiencies are not limited to technical shortcomings. The ever changing landscape of '*lean*' enterprise software development requires engineers to be equipped with abilities beyond the technical. How can higher education help students become great software developers in this context? As a first step towards answering this question, we present the results of a systematic literature review, focusing on *noncognitive abilities*, better known as '*soft skills*'. Our results identify self-reflection, conflict resolution, communication, and teamwork as the top four taught skills. Internships and capstone projects require more attention as a teaching aspect to facilitate the learning of multiple skills, including creativity. Interdisciplinary teaching and group composition are other important factors that influence learning. By providing novel insights on relationships between noncognitive abilities and teaching aspects, this work contributes to the continuous improvement of software engineering curricula. These findings may also serve as a springboard for further investigation of certain undervalued skills.

## 1. INTRODUCTION

When teaching aspiring software developers, educators are faced with the question: '*What makes a software engineer stand out in his or her profession?*'. Possible answers might include the ease of coming up with sound technical solutions, or the empathic ability to work well with others. Technical proficiency used to be the primary condition for success [1], but this knowledge is no longer enough [2]. Researchers, educators, and practitioners have all tried to answer the question what makes modern developers stand out. Papers ranging from 1994 [3] to 2018 [4] share a global message that still does not seem to be fully carried out by higher education. There is still no general consensus reached.

In this paper, we perform a *systematic literature review* to gain a deeper understanding of how modern engineering education has been shaped towards this new skillset. The formalized process of a systematic review has proven to be very insightful for identifying the current state-of-the-art on a given subject, and has been used widely in different fields, including software engineering research [5]. We focus our review on software engineering education, aiming to answer the following research question:

> ***What is the current state-of-the-art of teaching noncognitive abilities in software engineering education?***

The remainder of this paper is divided into the following sections. Section 2 describes background information on noncognitive skills and abilities, and why they are of growing importance, including related work on this topic. Section 3 clarifies the systematic review process we have used. Next, in section 4, we present and discuss the results of the review. Possible threats to validity are identified in section 5, while the last section, part 6, concludes this work.

## 2. BACKGROUND AND RELATED WORK

Defining boundaries for the term '*noncognitive abilities*' is becoming increasingly hard as different authors interpret it differently [6]. We have found other frequently used terms that slightly differ in meaning, although they have been used as synonyms in the literature. These terms range from *soft skills*, *21st century skills*, *intangible skills*, *human factors*, *interpersonal skills* and *generic competencies* to *social & emotional intelligence* and *people skills*. Multiple interpretations make the comparison of papers quite difficult. For the purpose of this literature review, we used all these different synonyms in our search, in order to obtain a broad picture of the domain.

A large amount of research on soft skills for software engineers exists, including specific industry studies [2,7]. Most of these studies do not explicitly focus on the education system itself. Instead, they highlight shortcomings from the point of view of the industry with the help of e.g. job ad analysis and focus groups. For instance, the SWEBOS (*Software Engineering Body of Skills*) framework by Sedelmaier, et al. [2] highlights the shortcomings of soft skill inclusions in the conventional SWEBOK (*Software Engineering Body of Knowledge*) model [1].

Examples of recent literature reviews similar to ours are [8] by Garousi, et al. in 2018 and [9] by Radermacher, et al. in 2013. Although these works provide insight into required noncognitive abilities, they do not focus solely on education, as our research does. Other publications delve deeper into specific subjects, such as the impact of pair programming [10,11], and the added benefit of improved confidence and self-esteem. Lenberg, et al. conducted an interdisciplinary research of '*the psychology of programming*', redefining and reviewing behavioral software engineering [12]. This combination of practical psychology with software engineering yields promising results for understanding what makes great developers tick [13].

Another approach to identify skills is by investigating success stories in software development. Dutra, et al. explored high performance teams using a systematic literature review [14], while Li, et al. simply asked practitioners: '*what makes a great software engineer?*' [15] Unsurprisingly, more than 50% of the answers can be categorized as non-technical, attributed to external (teammates) and internal (personal characteristics) factors.

These publications all strongly indicate the need for a revision in software engineering education, beyond technical knowledge. However, academic knowledge and skill requirements do not always perfectly match the abilities required from a software developer in the industry. Radermacher, et al. use the term '*knowledge deficiency*' to describe this lack of skills [9]. It seems that these deficiencies are given little attention but are becoming more and more important in the industry because of the way software is created: together, in close collaboration [4,6]. By providing a literature overview on noncognitive skills in software engineering education, we identify the current state of knowledge on teaching noncognitive skills to future software developers.

## 3. METHODOLOGY

The Systematic Literature Review (SLR) procedure we followed is an adaptation of Kitchenham's guidelines, which was adapted specifically for software engineering [5,16]. After establishing a review protocol, we formulated two research questions to steer the reviewing process. These questions, together with the search strings and criteria, helped us narrow down the publication result list, filtering out irrelevant papers.

The following research questions were identified:

- **Q1**: *Which noncognitive abilities have been identified by educators as important to teach software engineering students?*
- **Q2**: *How have those abilities been successfully taught?*

The ACM Digital Library and IEEE Xplore libraries were used as our main search services as they provide export functionalities, and our institution provides full-text access. Our focus is on software (1) engineering/development (2) noncognitive skills (3) in education (4). These are the search strings used to gather data, in conjunction with the many synonyms of '*noncognitive skills*' as described in section '2: `"software"` (1) AND (`"engineering"` OR `"developer"` OR `"development"`) (2) AND [synonym] (3) AND (`"education"` OR `"educational"` OR `"teaching"` OR `"curriculum"`) (4).

Additional ad-hoc searching via index aggregation services such as Google Scholar was needed to make sure we did not miss any major work. A technique called '*snowballing*' was adopted to consider publications from reference lists of papers in the review pool [16]. As part of the quality control, papers were required to contain at least some empirical evidence. Papers written in languages other than English were not taken into consideration. Also, to keep the results relevant for modern software engineering and to further limit the amount of results, publications older than 2014 were not included. However, this date limitation has not been applied while ad-hoc searching.

There is a lot of existing literature about engineering education in general. However, the software engineering discipline is unique compared to other engineering disciplines because of the complete absence of a fabrication cost and the increased speed of innovation [17]. This could result in different requirements of non-technical skills for each field. Therefore, papers in this review will not be included without the explicit mention of software. Also, to be able to answer question 2, we are only interested in success stories, thereby eliminating negative results.

1962 publications were initially screened based on their title, keeping 146 results. The next screening phase was based on paper abstracts, keeping 60 results. The last screening phase was based on the entire publication content, evaluating quality and applicability. In the end, 26 papers remained to be discussed in section 4. The complete dataset of all considered publications including extracted data can be found at https://people.cs.kuleuven.be/~wouter.groeneveld/slr/.

## 4. RESULTS AND DISCUSSION

A lot of different approaches towards integrating noncognitive skills into the curriculum have been found. These approaches maintain different time frames, ranging from one-day projects to extensive capstone projects and internships. The papers include diverse research methods, from single case studies to literature reviews [12]. The combined dataset has been published from 16 countries world-wide, with Germany (8 papers), Israel (3), and USA (3) on top. The advantage of our systematic review is that this combined data provides stronger evidence than each individual study.

Some papers have a narrow focus, targeting only a single skill [18,19] or a single teaching aspect [20–22]. Others are fairly broad, discussing soft skills in general [8,23]. We will discuss the results for literature review questions (Q1, Q2) individually, concluding with connections between the two.

### 4.1. Which skills are perceived as important? (Q1)

*Table 1: Identified skills.*

| Key | Skill | #papers |
|---|---|---|
| S01_Comm | Communication | 25 |
| S02_Team | Teamwork/dynamics | 25 |
| S03_Refl | Self-reflection | 13 |
| S04_Conf | Conflict resolution | 13 |
| S05_Mntr | Mentoring | 10 |
| S06_Ledr | Leadership | 7 |
| S07_Moti | Motivation | 6 |
| S08_Role | Role awareness | 4 |
| S09_Cult | Cultural Intelligence | 4 |
| S10_Crea | Creativity | 4 |
| S11_Ethi | Ethics | 3 |
| S12_Lifl | Lifelong Learning | 3 |
| S13_Empt | Empathy | 2 |

Table 1 contains a list of extracted non-technical skills, identified as important to teach software engineering students. Due to the vague definitions of each term, interpretations might overlap. Elaborate descriptions of the skills were mostly absent in reviewed publications, making it difficult to generalize or group results. We have refrained from using our own interpretation of these terms and only marked a term as present in a certain publication if it appears literally. Also, the absence of a term does not mean it is not deemed as important to teach by the authors, since it may yet be included implicitly in the program.

The reported non-technical skills in SWEBOS correspond roughly to the results in Table 1. SWEBOS uses the following skill groups: collaboration with others, communication, structuring one's way of working, personal competencies, consciousness of problems, problem solving, and further competencies [2]. It is difficult to say whether competencies from SWEBOS, such as '*acceptance of responsibilities*', can be seen as a combination of *role awareness*, *motivation*, and

*leadership*. The same applies to '*handling criticism*' or '*working calmly and efficiently under stress*': they show overlapping but cannot be identified with a single result. As a consequence, the SWEBOS list is not directly visible in Table 1. *Time management* and *problem solving* are not part of our interpretation of the term 'noncognitive skills'.

**Communication** and **teamwork** were the most common identified skills, and have also been the most commonly identified in industry surveys [9]. We did not make the distinction between written and oral communication, such as presentation skills. Holzer, et al. advocate for a separate course devoted to communication [18] while others integrate it more implicitly into the curriculum [24–26]. There is a clear correlation between *communication* and *teamwork*: when one skill appears in a paper, the other also occurs. The term '*teamwork*' is also very common within software engineering education, as students usually need to finish at least one form of project within a team during the program.

**Mentoring**, and **being mentored**, has been identified as a skill for both students and teachers. Students can act as an 'advisor' (*mentor*) in special programs [27], or can be mentored by peers or the teaching staff. Most papers left room for mentoring as part of the (capstone) project. Most mentoring happens outside of classrooms, such as the 25% time spent as part of the 'Communications & Networks' course design outlined by Cukierman, et al. [23].

**Conflict resolution** also appears in conjunction with *teamwork* and *communication*. Most papers view this skill as the classic interpersonal mediation skill when working together on capstone projects [20]. However, some papers introduced *conflict resolution* as an intrapersonal skill when developing your own career [28] or thinking about global issues introduced in a communications course [18].

**Leadership** suggests taking on a leading role in student team projects [20,29] or group discussions [18]. It can also imply spontaneously taking on the role of mentor when a fellow student is in need of help. There are clear connections between *leadership* and *teamwork*: being a good leader demonstrates the ability to work well within a team.

**Self-reflection** comes in many forms, ranging from general self-improvement [27,30] to specific reflections on the skills learned in the form of surveys [22,26] or assessment tools [21]. The better the student's ability to reflect, the better the ability to absorb other skills. Ebentheuer, et al. reported on a soft skill guidance program that ran successfully for years at their faculty, in which *self-reflection* plays a central role [27]. It is also important when working with an interdisciplinary group of students [24], or when thinking about one's future role as a software engineer in society [28]. Educators acknowledge the importance of *self-reflection*: we found the term in 50% of our results. It can be seen as the main enabling skill that increases the likelihood of learning anything else:

> *"Self-reflection is a crucial enabler for self-improvement in all areas of life."* [30]

**Motivation** as a separate skill denotes the importance of being driven to learn new skills. Students are likely to be more motivated when consistently working together [31]. 4 out of 6 occurrences of *motivation* also included *self-reflection*.

**Role awareness** also requires some *self-reflection* to see how a software engineer can play a meaningful role in our modern society [2,32]. Acheson, et al. advocate for a deeper understanding of specific strengths in different engineering roles [28].

***Ethics*** appears in conjunction with *role awareness* in the work of Li et al. [32]. Ethics of software engineering is a topic that usually appears in courses such as 'Soft Concepts of Computer Science' introduced by Hazzan, et al. [33].

***Cultural Intelligence/Diversity*** has been explicitly mentioned in 4 papers. It is a critical skill for future developers as engineering teams can be culturally diverse. This is especially the case with global software development.

***Empathy*** is closely related to *cultural diversity* and *ethics*, but there was no overlap found in the usage of these terms. Levy is the only author to completely focus on *empathy* in an interdisciplinary course [21].

***Creativity*** scores surprisingly low at only 4 occurrences. It is mostly related to open assignments in project-oriented learning [22,27,34,35]. We firmly believe that *creativity* is important to arrive at a good software solution, although hardly any explicit attention is paid to it.

***Lifelong Learning*** is the odd one out among the identified skills. It describes a set of skills, such as creativity, leadership and problem solving in general. Lifelong learning is an attitude, not an individual skill. However, since it was explicitly mentioned in several papers, we decided to include it in the results.

### 4.2. How have these skills been successfully taught? (Q2)

Table 2: Identified levels at which to integrate skills into the curriculum.

| Key | Level | #papers |
|---|---|---|
| L1_Course | Lectures of a single course | 13 |
| L2_Projec | Projects within a course | 9 |
| L3_Curric | Throughout entire curriculum | 5 |
| L4_Capsto | Capstone projects | 4 |
| L5_Intern | Internships | 2 |
| L6_Module | Modules within a course | 1 |

Table 3: Identified important teaching aspects.

| Key | Teaching aspect | #papers |
|---|---|---|
| A1_Inter | Interdisciplinarity | 7 |
| A2_Group | Group composition | 5 |
| A3_Colla | Collaborative Tools | 4 |
| A4_Testi | Assessment Tools | 4 |
| A5_Activ | Active Learning | 3 |
| A6_Video | Video Watching | 2 |
| A7_Playf | Playful Learning | 2 |

The skills identified in Table 1 can be taught in different ways. Table 2 shows the different levels at which the included publications try to integrate these skills into the curriculum. Table 3 contains a list of the different aspects of teaching on which these publications focus. Both topics are equally relevant for research question two, and will be discussed below, starting with levels of integration.

***Lectures*** (13 appearances) as part of a specific course and ***Projects*** (9) are the most popular levels. These have always been the classic tools for introducing new goals in the curriculum. Project courses can be converted completely into 'service-learning projects', to contribute to the community and further practice soft skills [28]. These projects involve working on real-world problems beyond university boundaries. Service-learning projects also strengthen the bond between community and the university.

***Curriculum-wide*** incorporation is the ultimate way to integrate noncognitive skills into every facet of the whole software engineering program. 'Soft skills must be included in curricula', according to Garousi, et al. [8]. Sedelmaier, et al. propose SWEBOS to guide curriculum changes, instead of looking at SWEBOK [2].

**Capstone projects** and **Internships** might require more time than a few seminars, but according to [20], the effort is worth it in the increased amount of skills learned. *Capstone* and *internship* projects pay off even more if they are real projects developed in-company instead of at the university [36]. These projects might not lead to students completely mastering skills such as *teamwork* and *conflict resolution*, but they will at least learn the relevance of these skills:

> *"We designed the course so as its main learning outcome is that students internalize how relevant it is having and developing critical soft skills to succeed in software development projects." [36]*

**Modules within a course**, such as *interdisciplinary* seminars, indicate the integration of specific parts in an existing course, without completely redesigning it. For instance, Acheson, et al. invite industry experts for seminars in their career development course. This integration is a good alternative, because, as stated in [28]:

> *"1) does not require much additional classroom time and instructor efforts; 2) can be seamlessly integrated into existing course materials; and 3) can start from the student's freshmen year and continues through their undergraduate study in the program."*

**Interdisciplinarity**, as the most important teaching aspect (7 appearances), refers to the mixing of teaching staff and students between different faculties, but also between industry and academia. Chatley, et al. introduce industry-relevant content by inviting guest speakers that talk about real-world problems [37]. In the communications module of Holzer, et al., the staff of the Technology faculty worked together with the Social and Human Science faculty [18]. Vicente, et al. sent students divided into *interdisciplinary* teams across programs on a 3-day team-building event to reinforce team spirit [38]. This cross pollination has proven to be effective to teach *empathy*, *cultural diversity* and *ethics*.

**Group composition** is another major factor in enhancing teaching of *mentoring*, *conflict resolution* and *leadership* [39]. In particular, cohesion within a student team has shown to influence *motivation*, productivity and performance [31]. Teaching staff might assemble teams themselves, as was the case with the *interdisciplinary* teams of [38]. Alternatively, surveys may be used to group students with the same primary goal, strengthening certain competencies within the group [30].

**Collaborative tools** have proven to be effective in assisting the skill learning process. Papers reported on the usage of digital tools such as forums and social media to facilitate learning communication [23,40]. However, these tools can just as well be simple analog post-it notes when employing agile practices in a project [37], or when reflecting upon the learned knowledge [30].

**Assessment tools** have to be developed to evaluate students' abilities during and at the end of a course [41]. In comparison to hard skills, soft skills are difficult to pin down in terms of grading. The speed of evaluation matters: employing a '*fast feedback cycle*' allows students to practice their reflection skill more often [37].

**Active Learning** and **Video Watching** help foster further learning of soft skills. Galster, et al. opted for what they call '*Active Video Watching*', integrating interactive activities into videos to reduce the resource costs involved in teaching [25]. Videos are also used as supportive material together with classes [18]. Especially for soft skills, active engagement with others is required to construct mental models. This interaction itself again requires the application of soft skills [33].

***Playful learning*** has been used in the form of gamification [34] and experimental investigations to discover unknown content [22]. Learning through play with supportive guidance creates more space to discover, improvise and challenge. Therefore, this method directly influences the *creativity* and *motivational* skills. *Playful learning*, however, is apparently still in an early stage of adoption in software engineering education, as witnessed by the fact that this term appeared only in 2 of our papers.

### 4.3. Relationships between results

Figure 1 visualizes the relationship between identified noncognitive skills and topics relevant for teaching these skills. To put emphasis on skills, contents from Table 2 and 3 have been combined in a single axis, totaling 13 topics. The following interesting connections have been discovered by interpreting the visual links between skill and teaching topic, or the striking absence of a link where we would expect one. These findings are even more easy to deduce from the interactive version of this diagram.

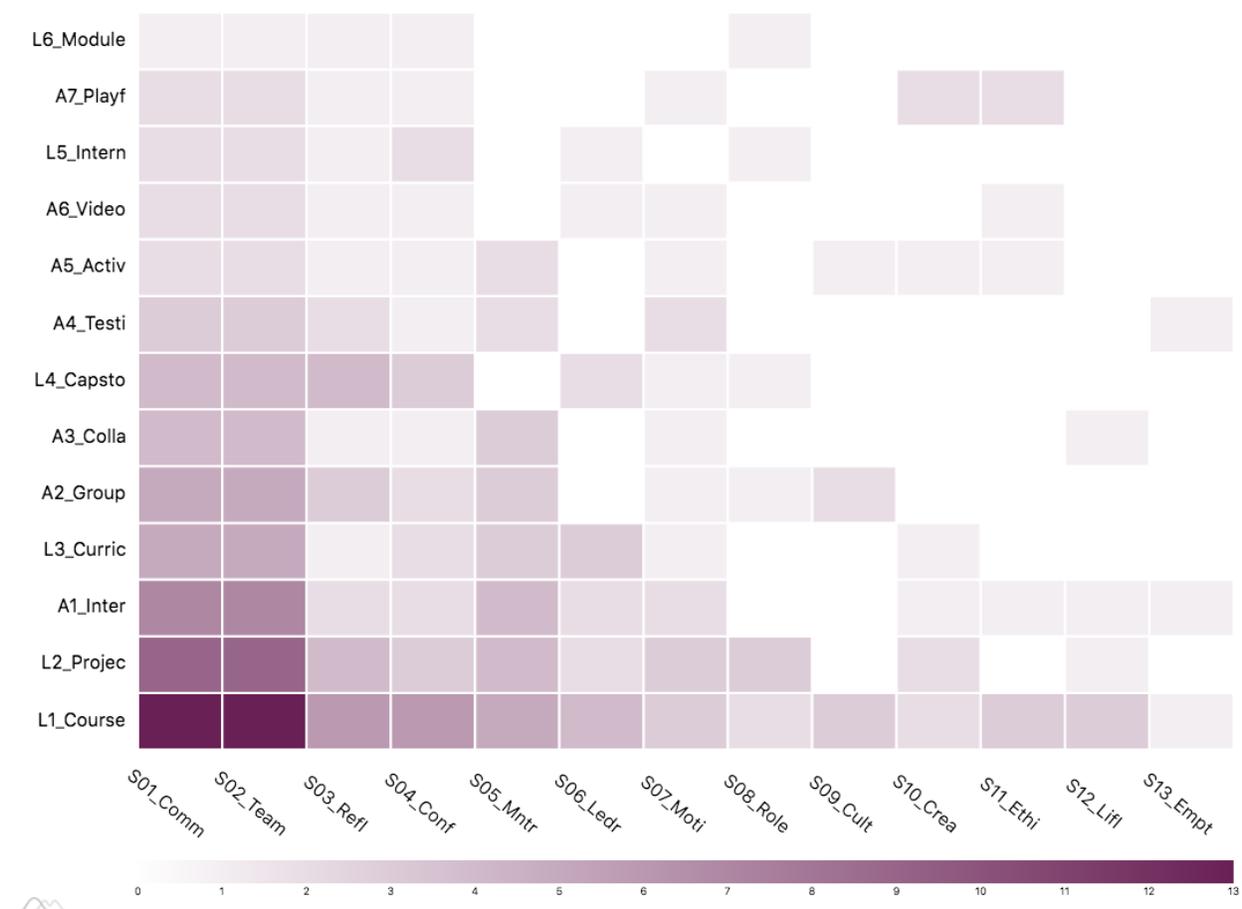

*Figure 1: An overview of the relationships between skill, aspect and level, outlined as a heatmap. Keys can be translated into corresponding values via tables 1, 2, and 3. An interactive visualization is available at https://people.cs.kuleuven.be/~wouter.groeneveld/slr/.*

**Generally popular combinations**: *self-reflection* and *conflict resolution* each appear in 13 out of 26 papers (50%) included in this study. Each time either is mentioned, all identified teaching aspects are also mentioned. The same effect holds for the two most common skills, *communication* and *teamwork*. These four abilities are the top studied skills in publications. *Motivation* is the next highly linked skill, missing only *internships* and *modules within a course*. It surprises us that *capstone projects* are not linked with *mentoring* (or *being mentored*), as one would expect that this kind of guidance is crucial to complete such a project.

***Collaborative Tooling* and *Test Assessment* show promise**: While *collaborative tools* assist the learning of *communication* skills, they are not yet used to facilitate the teaching of *ethics*, *empathy,* and *cultural diversity*. Forums and social media could also be deployed for these abilities. Testing students' behavior on non-technical abilities also shows encouraging results. We believe this can be further extended by involving *role awareness*, *ethics,* and *leadership*. *Creativity* is much more difficult to assess using tests.

***Internships* and *capstone projects* seem underused**: Besides the four top studied skills, *internships* are only connected to *leadership* and *role awareness*. *Capstone projects* are also linked to *motivation*. The lack of more links is surprising considering the outcry to bring industry and academia closer together. Strangely enough, there is only a very weak link between *leadership* and *internships*, and none at all between *mentoring* and *capstone projects*.

***Interdisciplinarity* is an advantage in teaching beyond the technical**: While *interdisciplinary* teaching aspects are only mentioned in 7 out of 26 papers (27%), they do cover all skills except *role awareness* and *cultural diversity*: 11 out of 13 skills (85%). The use of an interfaculty team, or even an *interdisciplinary* team across industry and academia, is strongly recommended [18].

***Lifelong learning* is perceived as a secondary goal**: The low number of appearances of *lifelong learning* (3), combined with the low number of coupled teaching aspects and levels (4 out of 13, 30%), leads us to conclude that teaching students the importance of *continuous training* is considered only a secondary or implicit goal. Perhaps educators feel that it is being (semi-)automatically induced by other skills. We believe it should instead be given the greatest attention, especially in an ever-changing world like software engineering. Figure 1 confirms the relationship between *lifelong learning* and *self-motivation* or *self-reflection*.

***Creativity* is absent in bigger project development**: *Creativity* appears in 4 out of 26 papers (15%). While it is related to a reasonable number of teaching aspects and levels (6 out of 13, 46%), it is not explicitly found when students embark on bigger projects such as *capstone projects* and *internships*. The papers included in our study never focus explicitly on *creativity* alone.

***Lectures* might not be the best way to induce noncognitive skills**: It is interesting to note that lectures, appearing 13 times (50%) and connecting with all skills, also seem to be preferred to induce more practical skills. Perhaps this is simply because it is a well-established method to teach theoretical knowledge. One could ask whether this is the most effective way to engage students.

## 5. THREATS TO VALIDITY

A possible threat to the correctness of our results is that the list of non-technical skills we identified in Table 1, or the teaching aspects and levels in Table 2 and 3, is incorrect or misaligned, and that some of these concepts are misinterpreted. Even though we recognize this possibility, we consider it unlikely, given the used methodology which reduces the risk of making these errors.

Limited visibility of publications may have led us to exclude certain important work. Most papers focus on soft skills in general, but some are more devoted towards certain individual skills. This will influence the visualization of the relationships between skills and aspects. Since the data extraction process is a manual process, we cannot guarantee that some papers, skills or aspects, have not mistakenly been excluded or missed. Therefore, we discuss our results as a whole and tried not to draw conclusions based on a single paper or identified relationship.

## 6. CONCLUSION

The results of our systematic literature review based on 26 papers identify which noncognitive abilities are perceived as important by educators, and how these are currently being taught, i.e. at which level they are integrated in the curriculum and to which aspects attention is being paid. We discussed each skill, aspect, and integration level individually to provide some context, highlighting the most and least common occurrences. By looking at the relationship between skill and aspect with the help of Figure 1, we discovered popular combinations and interesting trends in software engineering education.

It is clear to us that collaborating across academia and industry has had a major positive impact on the teaching of non-technical abilities. The first steps have already been taken to successfully blend practical psychology and philosophy with software engineering, but there is still room for improvement, both in depth and breadth.

Our findings may serve as a foundation to further investigate how to integrate the teaching of noncognitive skills into the curricula. For instance, this work can be compared to findings from industry surveys, further investigating certain skills or teaching methods based on the greatest common denominator. This will be our next step in contributing to research of soft skills in software engineering education.

Based on the results and conclusions of this study, we reckon that the following steps should be taken to strengthen the current software engineering curricula. First, the program should focus more on interdisciplinary teaching, not only by inviting lecturers from other faculties, but also from outside the university. Next, noncognitive abilities should be examined in more detail in combination with external internships and capstone projects. Lastly, skills such as creativity and a strong emphasis on lifelong learning should be induced in all available courses, including technical ones.

## 7. REFERENCES

[1] P. Bourque, R. E. Fairley, and others, (2014), Guide to the software engineering body of knowledge (SWEBOK (r)): Version 3.0, IEEE Computer Society Press.

[2] Y. Sedelmaier and D. Landes (2014), Software engineering body of skills (SWEBOS), in Global engineering education conference (EDUCON), 2014 IEEE, pp. 395–401.

[3] B. Friedman and P. H. Kahn (1994), Educating computer scientists: Linking the social and the technical, Commun. ACM, vol. 37, no. 1, pp. 64–70.

[4] L. F. Capretz and F. Ahmed (2018), A call to promote soft skills in software engineering, arXiv preprint arXiv:1901.01819.

[5] Z. Stapić, E. G. López, A. G. Cabot, L. de Marcos Ortega, and V. Strahonja (2012), Performing systematic literature review in software engineering, in CECIIS 2012-23rd international conference.

[6] G. Matturro, F. Raschetti, and C. Fontán (2015), Soft skills in software development teams: A survey of the points of view of team leaders and team members, in Proceedings of the eighth international workshop on cooperative and human aspects of software engineering, pp. 101–104.

[7] M. Stevens and R. Norman (2016), Industry expectations of soft skills in it graduates: A regional survey, in Proceedings of the Australasian Computer Science Week Multiconference, p. 13. ACM.

[8] V. Garousi, G. Giray, E. Tüzün, C. Catal, and M. Felderer (2018), Closing the gap between software engineering education and industrial needs, arXiv preprint arXiv:1812.01954.

[9] A. Radermacher and G. Walia (2013), Gaps between industry expectations and the abilities of graduates, in Proceeding of the 44th ACM technical symposium on computer science education, pp. 525–530.

[10] M. Schumm, S. Joseph, I. Schroll-Decker, M. Niemetz, and J. Mottok (2012), Required competences in software engineering: Pair programming as an instrument for facilitating life-long learning, in Interactive collaborative learning (ICL), 2012 15th international conference on, pp. 1–5.

[11] K. Chen and A. Rea (2018), Do pair programming approaches transcend coding? Measuring agile attitudes in diverse information systems courses, Journal of Information Systems Education, vol. 29, no. 2, pp. 53–64.

[12] P. Lenberg, R. Feldt, and L. G. Wallgren (2015), Behavioral software engineering: A definition and systematic literature review, Journal of Systems and software, vol. 107, pp. 15–37.

[13] D. Graziotin, X. Wang, and P. Abrahamsson (2015), Understanding the affect of developers: Theoretical background and guidelines for psychoempirical software engineering, in Proceedings of the 7th international workshop on social software engineering, pp. 25–32.

[14] A. C. Dutra, R. Prikladnicki, and C. França (2015), What do we know about high performance teams in software engineering? Results from a systematic literature review, in Software engineering and advanced applications (SEAA), 41st euromicro conference on, pp. 183–190.

[15] P. L. Li, A. J. Ko, and J. Zhu (2015), What makes a great software engineer? in Proceedings of the 37th international conference on software engineering-volume 1, pp. 700–710.

[16] C. Wohlin (2014), Guidelines for snowballing in systematic literature studies and a replication in software engineering, in Proceedings of the 18th international conference on evaluation and assessment in software engineering, p. 38.

[17] M. Young and S. Faulk (2010), Sharing what we know about software engineering, in Proceedings of the fse/sdp workshop on future of software engineering research, pp.439–442.

[18] A. Holzer, S. Bendahan, I. V. Cardia, and D. Gillet (2014), Early awareness of global issues and development of soft skills in engineering education: An interdisciplinary approach to communication, Information technology based higher education and training (ITHET), pp.1–6.

[19] M. Levy (2018), Educating for empathy in software engineering course, in Joint proceedings of refsq-2018 workshops, doctoral symposium, live studies track, and poster track.

[20] G. Zheng, C. Zhang, and L. Li (2015), Practicing and evaluating soft skills in it capstone projects, in Proceedings of the 16th annual conference on information technology education, pp. 109–113.

[21] M. Marques, S. F. Ochoa, M. C. Bastarrica, and F. J. Gutierrez (2018), Enhancing the student learning experience in software engineering project courses, IEEE Transactions on Education, vol. 61, no. 1, pp. 63–73.

[22] A. Soska, J. Mottok, and C. Wolff (2015), Playful learning in academic software engineering education, in Global engineering education conference (EDUCON), 2015 IEEE, pp. 324–332.

[23] U. R. Cukierman and J. M. Palmieri (2014), Soft skills in engineering education: A practical experience in an undergraduate course, in Interactive collaborative learning (ICL), 2014 international conference on, pp. 237–242.

[24] C. Andersson and D. Logofatu (2018), Using cultural heterogeneity to improve soft skills in engineering and computer science education, in Global engineering education conference (EDUCON), 2018 IEEE, pp. 191–195.

[25] M. Galster, A. Mitrovic, and M. Gordon (2018), Toward enhancing the training of software engineering students and professionals using active video watching, in Proceedings of the 40th international conference on software engineering: Software engineering education and training, pp. 5–8.

[26] Y. Sedelmaier and D. Landes (2014), Practicing soft skills in software engineering: A project-based didactical approach, in Overcoming challenges in software engineering education: Delivering non-technical knowledge and skills, IGI Global, pp. 161–179.

[27] A. W. Ebentheuer, J. Kammermann, and H.-G. Herzog (2017), Adveisor—analysis of an established soft skill program for students in the field of electrical engineering, in Global engineering education conference (EDUCON), 2017 IEEE, pp. 468–475.

[28] L. Acheson and R. Rybarczyk (2016), Integrating career development into computer science undergraduate curriculum, in Computer science & education (ICCSE), 2016 11th international conference on, pp. 177–181.

[29] F. Patacsil and C. L. S. Tablatin (2017), Exploring the importance of soft and hard skills as perceived by IT internship students and industry: A gap analysis, JOTSE, vol. 7, no. 3, pp. 347–368.

[30] V. Thurner, K. Schlierkamp, A. Böttcher, and D. Zehetmeier (2016), Integrated development of technical and base competencies: Fostering reflection skills in software engineers to be, in Global engineering education conference (EDUCON), 2016 IEEE, pp. 340–348.

[31] D. Tamayo Avila, W. Van Petegem, Y. Cruz Ochoa, and M. Noda Hernández (2018), A Correlational Study On Factors That Influence The Cohesion Of Software Engineering Students Teams, INTED2018 Proceedings, no. March, pp. 5697–5702, 2018.

[32] K. F. Li, J. Fagan, and I. Bourguiba (2016), Teaching professional practice and career development to graduate students, in Teaching, assessment, and learning for engineering (TALE), 2016 IEEE international conference on, pp. 398–402.

[33] O. Hazzan and G. Har-Shai (2013), Teaching computer science soft skills as soft concepts, in Proceeding of the 44th ACM technical symposium on computer science education, pp. 59–64.

[34] B. R. Maxim, S. Brunvand, and A. Decker (2017), Use of role-play and gamification in a

software project course, in 2017 IEEE frontiers in education conference (FIE), pp. 1–5.

[35]    H. Chassidim, D. Almog, and S. Mark (2018), Fostering soft skills in project-oriented learning within an agile atmosphere, European Journal of Engineering Education, vol. 43, no. 4, pp. 638–650.

[36]    M. C. Bastarrica, D. Perovich, and M. M. Samary (2017), What can students get from a software engineering capstone course?, Software engineering: Software engineering education and training track (ICSE-SEET), IEEE/ACM 39th international conference, pp.137–145.

[37]    R. Chatley and T. Field (2017), Lean learning-applying lean techniques to improve software engineering education, in Software engineering: Software engineering education and training track (ICSE-SEET), 2017 IEEE/ACM 39th international conference on, pp. 117–126.

[38]    A. Vicente, T. Tan, and A. Yu (2018), Collaborative approach in software engineering education: An interdisciplinary case, Journal of Information Technology Education: Innovations in Practice, vol. 17, no. 1, pp. 127–152.

[39]    A. Mujkanovic and A. Bollin (2016), Improving learning outcomes through systematic group reformation: The role of skills and personality in software engineering education, in Proceedings of the 9th international workshop on cooperative and human aspects of software engineering, pp. 97–103.

[40]    R. Y.-Y. Chan et al. (2017), Direct evidence of engineering students, in 2017 IEEE frontiers in education conference (FIE), pp. 1–5.

[41]    V. Thurner, A. Bottcher, and A. Kamper (2014), Identifying base competencies as prerequisites for software engineering education, in Global engineering education conference (EDUCON), 2014 IEEE, pp. 1069–1076.